\PassOptionsToPackage{small,bf}{caption}

\documentclass[smallextended]{svjour3}

\RequirePackage{fix-cm}
\usepackage{mathptmx} 
\smartqed
\journalname{Multimedia Tools and Application}
\usepackage{url}
\usepackage{epsfig}
\usepackage{latexsym}
\usepackage{microtype}
\usepackage{tabularx}
\usepackage{amsmath}
\usepackage{arydshln}
\usepackage{booktabs}
\usepackage{wrapfig}
\usepackage{pgfplots}
\usepackage{textcomp}
\usepackage{chngpage}
\usepackage{graphicx,adjustbox}
\usepackage{multirow, makecell}
\usepackage[group-separator={,}, group-four-digits=true, group-digits=integer]{siunitx}
\usepackage[notext,not1]{stix}
\usepackage{fancyvrb}
\usepackage{natbib}
\usepackage{tabularx}
\usepackage{colortbl}
\usepackage{xcolor}
\usepackage[T5,T1]{fontenc}
\pgfplotsset{compat=1.7}

\usepackage{hyperref}

\usepackage{tikz}
\newcommand*\circled[1]{\tikz[baseline=(char.base)]{
            \node[shape=circle,draw,inner sep=1pt] (char) {#1};}}
\usepackage[linguistics]{forest}

\let\citewopar\citet

\begin{document}
%Tam-Mal--dataset: A Gold-standard Dataset to Detect Sentiment and Offensive Content in Social Media for Low Resourced Malayalam Language ---- Dataset for Draviadian sentiment analysis and offensive language identification (DravidSentiOFF)
\title{ TrollsWithOpinion: A Dataset for Predicting Domain-specific Opinion Manipulation in Troll Memes
% Predicting domain-wise opinion manipulation in Troll Memes
%\thanks{The research work is supported in part by a research grant from Science Foundation Ireland (SFI) under Grant Number SFI/12/RC/2289$\_$P2 (Insight$\_$2), co-funded by the European Regional Development Fund as well as by the EU H2020 programme under grant  agreements 825182 (Prêt-à-LLOD), and Irish Research Council grant IRCLA/2017/129 (CARDAMOM-Comparative Deep Models of Language for Minority and Historical Languages)}
}
% Grants or other notes about the article that should go on the front
% page should be placed within the \thanks{} command in the title
% (and the %-sign in front of \thanks{} should be deleted)
%
% General acknowledgments should be placed at the end of the article.

%\subtitle{A classification dataset}

\titlerunning{Predicting domain-wise opinion manipulation in Troll Memes}        % if too long for running head

\author{Shardul~Suryawanshi\(^1\) \and
        Bharathi~Raja~Chakravarthi\(^1\) \and
        Mihael~Arcan\(^1\) \and
        Suzanne~Little\(^2\) \and
        Paul~Buitelaar\(^1\)
        %etc.
}

\authorrunning{Suryawanshi et al} % if too long for running head

\institute{Shardul~Suryawanshi \\
\email{shardul.suryawanshi@insight-centre.org} \\                      \and
         Bharathi~Raja~Chakravarthi \\ \email{bharathi.raja@insight-centre.org} \\ 
              \and
         Mihael~Arcan\\
         \email{mihael.arcan@insight-centre.org} \\
              \and
         Suzanne~Little\\ \email{suzanne.little@insight-centre.org} \\ 
              \\
         Paul~Buitelaar\\ \email{paul.buitelaar@insight-centre.org} \\
          \(^1\)Insight SFI Research Centre for Data Analytics, Data Science Institute, National University of Ireland Galway, Galway, Ireland \\
          \(^2\)Insight SFI Research Centre for Data Analytics, Dublin City University, Dublin, Ireland \\
}

\date{Received: date / Accepted: date}
% The correct dates will be entered by the editor
\maketitle
\begin{abstract}
Research into the classification of Image with Text (IWT) troll memes has recently become popular. Since the online community utilizes the refuge of memes to express themselves, there is an abundance of data in the form of memes. These memes have the potential to demean, harras, or bully targeted individuals. Moreover, the targeted individual could fall prey to opinion manipulation. To comprehend the use of memes in opinion manipulation, we define three specific domains (product, political or others) which we classify into troll or not-troll, with or without opinion manipulation. To enable this analysis, we enhanced an existing dataset by annotating the data with our defined classes, resulting in a dataset of 8,881 IWT or multimodal memes in the English language (TrollsWithOpinion dataset). We perform baseline experiments on the annotated dataset, and our result shows that existing state-of-the-art techniques could only reach a weighted-average F1-score of 0.37. This shows the need for a development of a specific technique to deal with multimodal troll memes.

% \PACS{PACS code1 \and PACS code2 \and more}
% \subclass{MSC code1 \and MSC code2 \and more}
\end{abstract}
\section{Introduction}
%\textcolor{red}{write about Sentiment analysis and offensive language identificaiton (more) https://link.springer.com/content/pdf/10.1007/s10579-020-09488-3.pdf}
Social media has given liberty in the form of anonymity to its users. People take advantage of the freedom of expression and express themselves under no censorship. Due to the lack of such a surveillance system, people fall prey to trolls or social media bullies. Troll is a self-centred egoist person who seeks the attention of the crowd by making demeaning remarks on an individual or a group \citewopar{tomaiuolo2020survey}. Supporters of trolling argue that it's about humour or freedom of speech. However, due to its subjectiveness, the abuse may cause great distress which varies from one individual to another. One such example emerged in the US state of Missouri in 2006, when 13-year-old Megan Meier took her own life after being bullied online \footnote{https://www.herts.police.uk/Information-and-services/Advice/Online-safety/Trolling-and-cyberbullying}.

Although their action (trolling) seems to intend humour most of the time, trolls may have an agenda in their mind that hides behind the humour. Based on the definition ``Trolling is the activity of posting a message via social media that tend to be offensive, provocative, or menacing to distract and digressive or off-topic content with the intent of provoking the audience \citewopar{hardaker2010trolling,bishop2014representations,mojica-de-la-vega-ng-2018-modeling}.''

Apart from spreading hate through the explicit use of profane language, trolls are also responsible for profound opinion manipulation \citewopar{coxall2013human} through their hidden agenda \citewopar{atanasov-etal-2019-predicting}. This hidden agenda might help an individual or a group to influence people positively or negatively. Troll farms played a significant role in outcomes of historical events such as the 2016 US presidential election and Brexit \citewopar{atanasov-etal-2019-predicting}. As these events shape the future of the world, we need a system that could identify trolls. The preventive measure could be used to monitor user behaviour and the pattern to identify the troll but by the time one identifies by such means the harm caused by the troll might be significant. Hence, we emphasize on identifying trolling with the help of a troll meme rather than troll (user).
Thanks to low network tariffs and smartphones now trolling can be done more creatively by the use of more effective media that can spread faster than a text i.e. memes. 

The term meme was coined by Richard Dawkins. According to him, ``it is a unit of cultural transmission or a unit of imitation and replication. Because it acts like genes and can self-replicate, mutate''. Trolls use a troll meme to spread hatred and to manipulate their audience. Image with text (IWT) memes are a popular choice by internet users; they have an image with text imposed on them \citewopar{du2020understanding}. One might miss the true meaning of such memes by considering just the text or image in isolation; hence both image and text should be considered. This makes the problem of identifying troll memes ``multimodal". We define troll memes as a meme,  which contains offensive text and non-offensive images, offensive images with non-offensive text, sarcastically offensive text with non-offensive images, or sarcastic images with offensive text to provoke, distract, and digressive or off-topic of the content with the intent to demean or offend particular people, group or race.

In this paper, we present the TrollsWithOpinion dataset, which is designed to address the real-world trolling problem in multimodal content (image+text). The dataset consists of 9000 memes( images+ text). We extract this dataset from the Memotion-analysis shared task \citewopar{chhavi2020memotion}. Memotion analysis shared task annotated the data for five different problems, namely, sentiment analysis, sarcasm detection,  offensive, motivation, and humour detection. We improve upon this by annotating for trolling content. Since memes are mostly used for trolling. We define the problem of identifying trolling in multimodal content at a different level.  As a result, we obtain a large dataset for identifying trolling on a different level in multimodal content.

\section{Related Work} 
%\subsection{Sentiment Analysis}

\paragraph{Trolling}: A lot of research has been done in the areas of hate speech detection \citewopar{schmidt-wiegand-2017-survey}, offensive speech detection \citewopar{zampieri-etal-2019-semeval}, and identifying trolling \citewopar{trac-2020-trolling} in text. Rather less emphasis has been put on other modalities such as image, audio or video. \citewopar{tomaiuolo2020survey} thoroughly assess the methods and challenges involved in identifying trolls in their ``A Survey on Troll Detection”. They highlight the use of post-based (based on the content in the online posts), thread-based (based on the analysis of thread of online posts), user-based (based on the overall attitude of the user) and community-based (based on the relationships of the user within the online community) methods to identify trolls. Our research is post-based which focuses on the identification of trolling (action) through the content of the online post. This technique tends to be quicker, and could thus prevent the harm that might be caused due to delay in identification of trolls using other methods mentioned earlier.

\paragraph{Opinion Manipulation}: Trolling can further lead to opinion manipulation; \citewopar{mihaylov-etal-2015-finding} and \citewopar{atanasov-etal-2019-predicting} point out the opinion manipulation caused due to trolling in news and political media.  
\citewopar{atanasov-etal-2019-predicting} propose an approach to identify left, right and centre (news feed) trolls using supervised and distantly supervised methods while \citewopar{mihaylov-etal-2015-finding} use an approach that distinguishes trolls from non-trolls based on the manual features derived from the number of comments posted, number of days in the forum, number of days with at least one comment, and number of publications commented on; both of these approaches are user-based. According to \citewopar{boatwright2018troll}, political trolls are state-sponsored agents who control a set of pseudonymous user accounts, also known as ``sock puppets” \citewopar{kumar2017army}, which dissipate misinformation \citewopar{karadzhov2018we} and propaganda with the purpose of swaying opinions, destabilizing society and influencing election outcomes. All of these aforementioned techniques emphasize text-based methods to identify trolls, while we concentrate on the multimodal aspect of the trolling in the form of a meme.

\paragraph{Memes} as a mode of trolling and opinion manipulation: Memes have become an integral part of online communication \citewopar{bauckhage2011insights}. Moreover, the manipulation of public opinion and propagation of hate among populations has become commonplace. \citewopar{zannettou2018origins} defines memes based on the virality of the post; a set of posts (image, video) can be referred to as a meme if those posts share a common theme and are disseminated by a large number of users. Zannettou’s (2018) social media analysis through the lens of memes gives an insight into the trend of sharing memes on mainstream (Reddit, Twitter) and fringe (4chan, gab) web communities. He proposed a pipeline based on a pHash and clustering algorithm, showing that the behaviour of sharing racist and/or political memes is more commonplace in fringe web communities. \citewopar{syuntyurenko2015network} suggests the possibility that a malicious bot can be disguised as a user; such a user can create a meme that might connect to a wider audience due to references to popular movies or other media. This might end up influencing the audience’s opinion or ideological orientation, which is also known as opinion manipulation.

\paragraph{Our contribution}: To the best of our knowledge, none of these studies has investigated trolling through memes which may or may not cause opinion manipulation. The research conducted by \citewopar{syuntyurenko2015network, boatwright2018troll, atanasov-etal-2019-predicting, mihaylov-etal-2015-finding} touches upon opinion manipulation caused by trolling, but does not lay out granular classes of opinion manipulation based on a targeted population. Our research aims to bridge this gap by providing definitions for granular classes of a troll or not-troll meme that may or may not cause opinion manipulation based on actions (Trolling with opinion manipulation, Trolling without opinion manipulation, Not-trolling with opinion manipulation, Not-trolling without opinion manipulation) that an internet user might engage in. Based on these actions, we categorise a meme into Troll\_opinion\_X, Not\_troll\_opinion\_X, Not\_troll\_without\_opinion, Troll\_without\_opinion where X represents targeted domain that could be either political, product or other. We define each meme class in Section \ref{Troll}, see below. Our TrollsWithOpinion dataset is based on these definitions, and we provide baseline results.

\section{Dataset} 
We collected data from memotion dataset and later enhanced the annotations as per our use case. The original dataset has memes in the form of image and OCR extracted text from Google vision API. Later, OCR extracted text has been verified and corrected by annotators. The annotated memes fall in Humorous, Sarcasm, Offensive, Motivation classes with quantified the intensity (not, slightly, mildly, very) to which a particular effect of a class is expressed, along with the overall sentiments (very negative, negative, neutral, positive, very positive). For our research, we only considered image and text associated with the image.

\subsection{Class Definitions} \label{Troll}

% \begin{figure*}[t]
%   \centering 
%   \includegraphics[width=1\linewidth,clip]{images/Images_for_troll.png} 
%   \caption{Examples for definition}
%     \label{fig:Example_pred} 
% \end{figure*}

% A troll meme is a post (image, audio, video, GIF) which intends to implicitly demean or offend an individual on the internet. Bishop et al. (xxxx) suggests that trolling is an activity whereby messages are posted on social media to offend or provoke other internet users. An internet user (troll) often operates to distract the audience with the intent of causing provocation. Dawkins (xxxx) coined the term ‘meme,’ stating that memes are a means of cultural transmission that can self-replicate and mutate. 

% Trolling leads to public opinion manipulation. This manipulation can give rise to negative effects for targeted individuals, sometimes even causing suicidal tendencies. One such notable case happened in the U.S. in 2006 when a 13-year-old, Megan Meier, took her own life after being bullied online. Some trolling is more subtle as showcased in events such as the 2016 U.S. presidential election and Brexit. In both cases, a large number of people were targeted and manipulated by troll farms to influence how they voted.

For the purposes of this research, we defined the activity of posting messages on the internet as trolling or not-trolling depending on whether manipulative intention is present and/or whether manipulative effect is present. Inspired from \citewopar{coxall2013human}, we define the following categories are used:

\paragraph{Trolling with opinion manipulation:} strategy or manoeuvre that makes offensive yet digressive comments to negatively influence opinion about the targeted individual or group (political party, company, minority, gendered group, religious group, etc.).

\paragraph{Trolling without opinion manipulation:} strategy or manoeuvre that makes offensive yet digressive comments about the targeted individual or group but does not influence opinion about the individual or group.

\paragraph{Not-trolling with opinion manipulation:} strategy or manoeuvre that makes a non-offensive or informative comment that intends to positively influence opinion about the targeted individual or group.

\paragraph{Not-trolling without opinion manipulation:} Strategy or manoeuvre that makes a non-offensive or informative comment about the targeted individual or group but does not intend to influence opinion about the individual or group.

\begin{figure*}[t]
  \centering 
  \includegraphics[width=1\linewidth,clip]{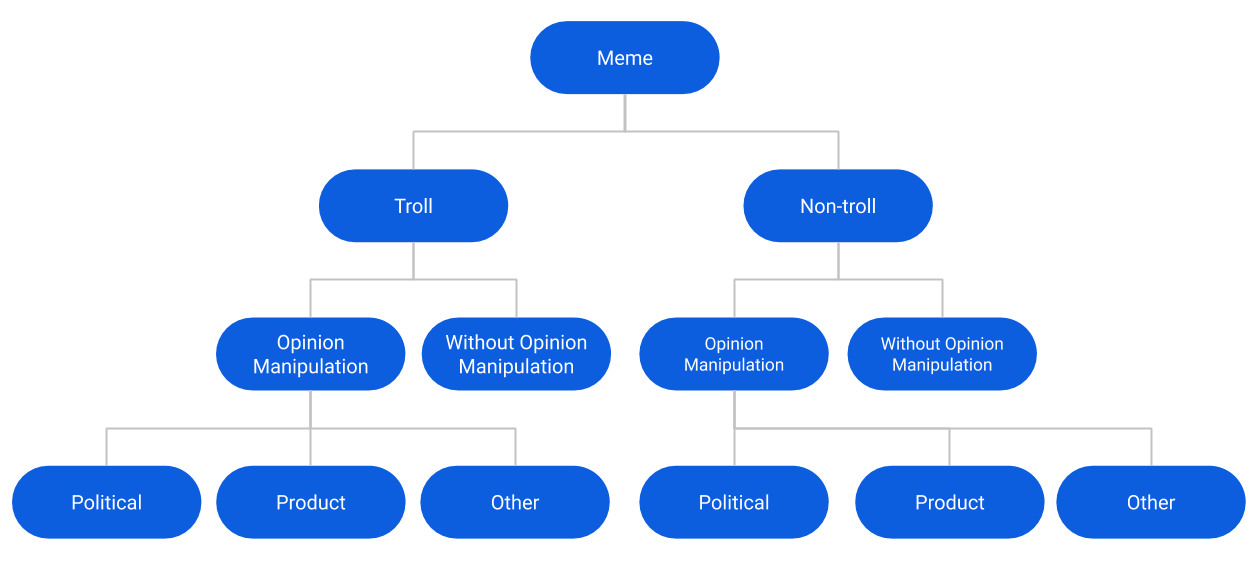} 
  \caption{Hierarchical division of memes into troll or not-troll based on the presence or absence of domain-wise opinion manipulation}
    \label{fig:class_hierarchy} 
\end{figure*}

Figure \ref{fig:class_hierarchy} shows the levels of classification. We define the classes in the form of hierarchy. On the first hierarchical level, we define multimodal or IWT meme into either troll or not-troll classes. On the second hierarchical level, we indicated if opinion manipulation is present (Opinion manipulation or Without opinion manipulation), and the third level represents the domain (political, product or other). In this work, we concentrated on the political and product domain by marking the rest as ``other".

\begin{figure*}[h]
  \centering 
  \includegraphics[width=1\linewidth,clip]{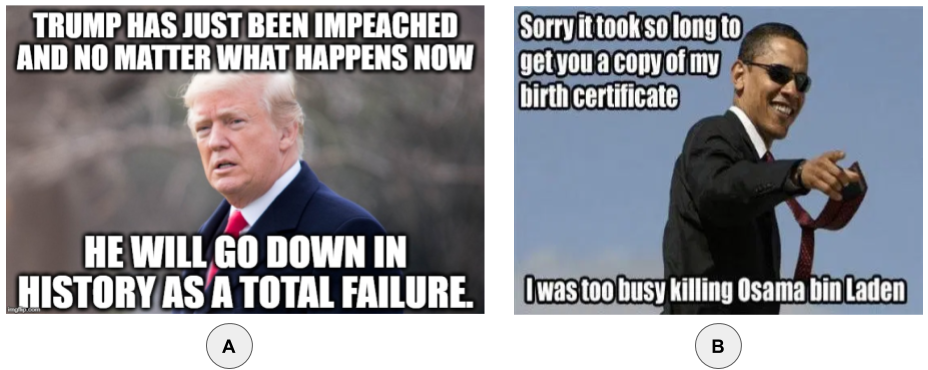} 
  \caption{Example on a troll and not-troll memes when political opinion manipulation is present}
    \label{fig:Example_political} 
\end{figure*}

Trolling is an action which can be predicted based on user history and the troll memes used. In this research, we are concerned with the identification of trolling based on troll memes, specifically memes that fall under an image with text (IWT) category. Troll IWT memes are characterized as having either one (either image or text) or both modalities being offensive. Based on the four actions defined above, we classified troll IWT memes into eight classes as below:

\paragraph{Troll\_opinion\_political}: This meme makes an offensive comment to negatively influence opinion about a political figure or party. An example circled \circled{A} from Figure \ref{fig:Example_political} is one such example, where a troll is criticizing Donald Trump in a demeaning manner.

\paragraph{Not\_troll\_opinion\_political}: This meme makes a non-offensive comment to positively influence opinion about a political figure or party. An example circled \circled{B} from Figure \ref{fig:Example_political} shows Barack Obama (44th president of the United States) smiling at the crowd with a caption stating the reason (killing Osama bin Laden) for not producing a birth certificate. This meme is trying to manipulate people into siding themselves with Obama since he killed a terrorist.

\begin{figure*}[h]
  \centering 
  \includegraphics[width=1\linewidth,clip]{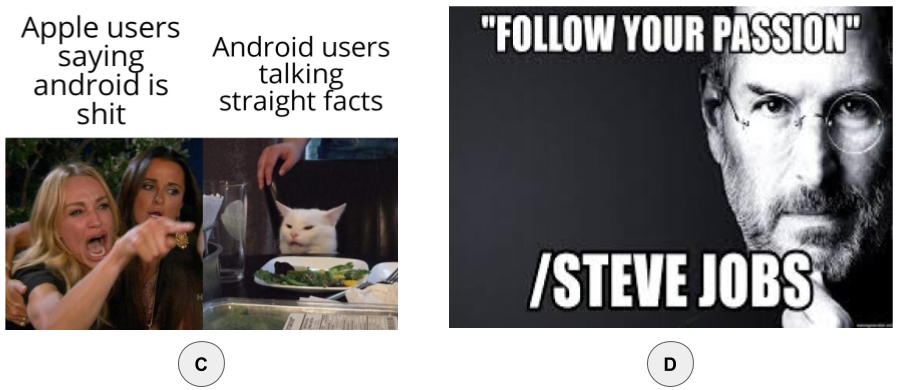} 
  \caption{Example on a troll and not-troll memes when product opinion manipulation is present}
    \label{fig:Example_product} 
\end{figure*}

\paragraph{Troll\_opinion\_product}: This meme makes an offensive comment to negatively influence opinion about a product (movies, video game, devices, daily household appliance, etc.) or individual that represents the product (e.g. company CEO, brand ambassador, etc.). An example circled \circled{C} from Figure \ref{fig:Example_product} is trolling Apple product users by stating supremacy of Android users over them. Implication of the meme could be a manipulation of new mobile users, as they might get inclined towards Android devices instead of Apple devices.

\paragraph{Not\_troll\_opinion\_product}: This meme makes a non-offensive comment to positively influence opinion about a product (movies, video game, devices, daily household appliance, etc.) or individual that represents the product (e.g. company CEO, brand ambassador, etc.). An example circled \circled{D} from Figure \ref{fig:Example_product} shows a picture of Steve Jobs (ex Apple CEO) with a motivational caption. As Steve Jobs was associated with Apple, this meme is trying to win over people by putting Steve Jobs in a spotlight (associating him with passion).

\begin{figure*}[h]
  \centering 
  \includegraphics[width=1\linewidth,clip]{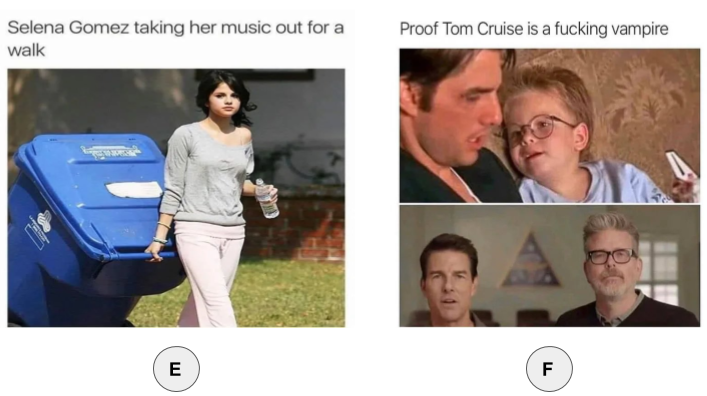} 
  \caption{Example on a troll and not-troll memes when opinion manipulation other than product and politics is present}
    \label{fig:Example_other} 
\end{figure*}

\paragraph{Troll\_opinion\_other}: This meme makes an offensive comment to negatively influence opinion about an individual or group based on their gender, ethnicity, sexual orientation, religious beliefs, etc. An example circled \circled{E} from Figure \ref{fig:Example_other} shows Selena Gomez (celebrity pop singer) with a trash bin, and the caption (Selena Gomez taking her music out for a walk) that goes along with the meme is meant for criticizing her music in a demeaning way. This manipulation might result in a drop in her popularity.

\paragraph{Not\_troll\_opinion\_other}: This meme makes a non-offensive comment to positively influence opinion about an individual or group based on their gender, ethnicity, sexual orientation, religious beliefs, etc. An example circled \circled{F} from Figure \ref{fig:Example_other} states that Tom Cruise (an American actor) is ageless by comparing him with a vampire. 

\begin{figure*}[h]
  \centering 
  \includegraphics[width=1\linewidth,clip]{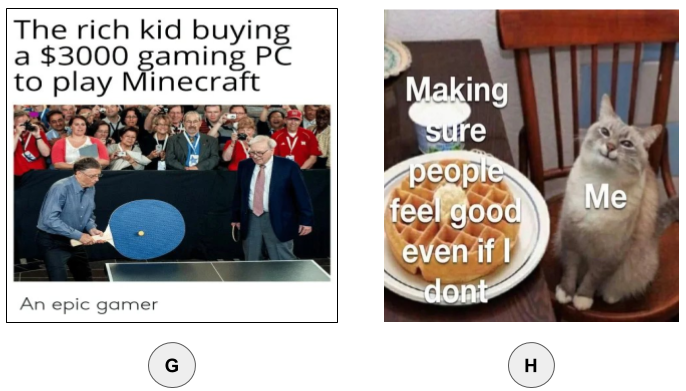} 
  \caption{Example on a troll and not-troll memes when opinion manipulation is not present}
    \label{fig:Example_not_troll} 
\end{figure*}

\paragraph{Troll\_without\_opinion}: This meme makes an offensive comment about the targeted individual or group but does not influence opinion about such individuals or groups. An example circled \circled{G} from Figure \ref{fig:Example_not_troll} is trolling people who play Minecraft (graphically less intensive game) on their high end gaming machines. But this meme is not changing opinion about the rich kid in a negative or positive way, hence it belongs to the Troll\_without\_opinion category.

\paragraph{Not\_troll\_without\_opinion}: This meme makes a non-offensive comment about an individual, group or product but does not influence opinion about them. An example circled \circled{H} from Figure \ref{fig:Example_not_troll} neither trolling nor changing opinion.

\begin{figure*}[h]
  \centering 
  \includegraphics[width=1\linewidth,clip]{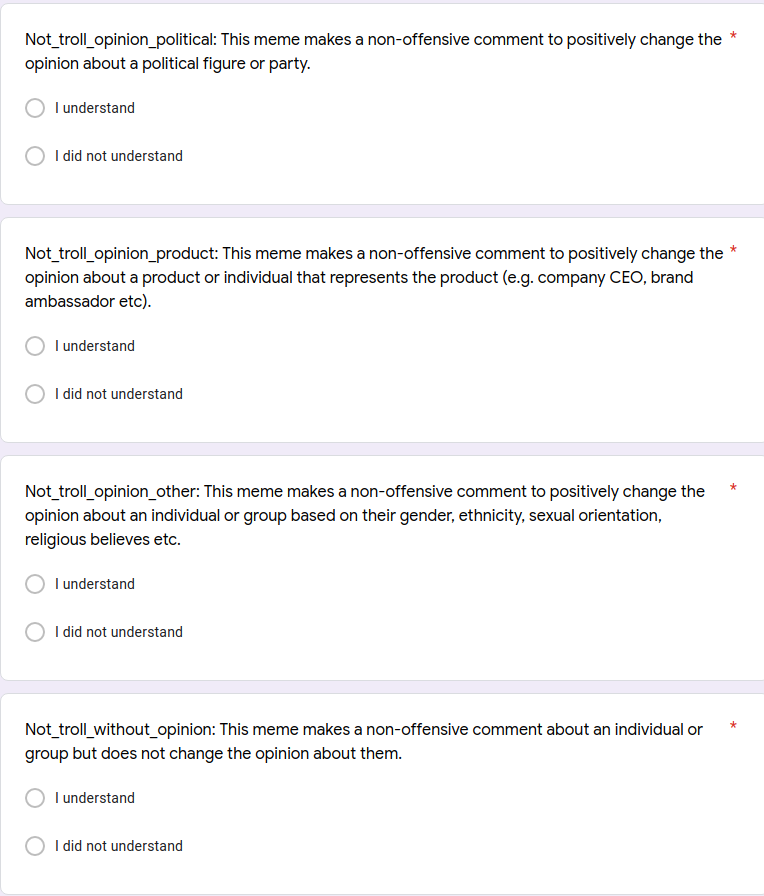} 
  \caption{A sample page from google form that explains not-troll meme classes}
    \label{fig:form_not_troll} 
\end{figure*}

\subsection{Annotation Process}
\begin{figure*}[t]
  \centering 
  \includegraphics[width=1\linewidth,clip]{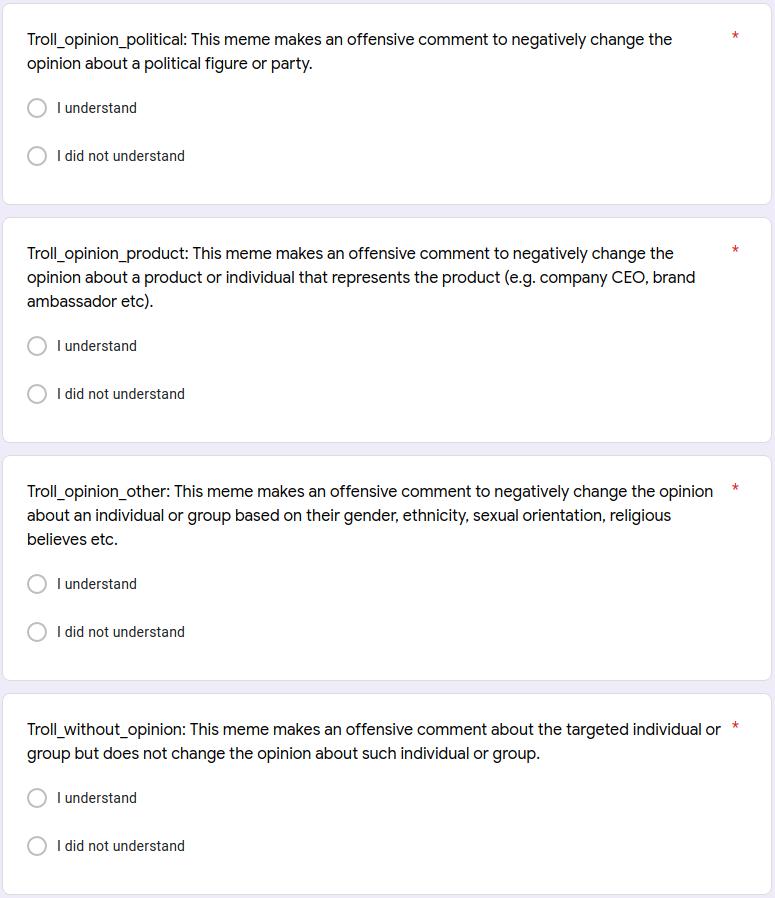} 
  \caption{A sample page from google form that explains troll meme classes}
    \label{fig:form_troll} 
\end{figure*}

% write about annotation with googel forms and write about the inter-annotator agreement.
% annotator statistics
We did annotation in two parts with the help of Google forms (examples could be found in Figure \ref{fig:form_not_troll}, Figure \ref{fig:form_troll}). Each google form consisted of 50 memes, with 10 memes on each page. Annotators (volunteers) were provided with the annotation guidelines, and it was made sure that they understood class definitions provided in the annotation guideline by analyzing their initial feedback. However, we also modified classes and definitions as per the feedback received from annotators in the pilot studies. To proceed with the annotation, annotators were instructed to agree on the terms, conditions and class definitions as shown in Figure \ref{fig:form_not_troll} and \ref{fig:form_not_troll}. Furthermore, they were allowed to discontinue at any point if they feel overwhelmed or offended due to the content present in forms.

\begin{figure*}[h]
  \centering 
  \includegraphics[width=1\linewidth,clip]{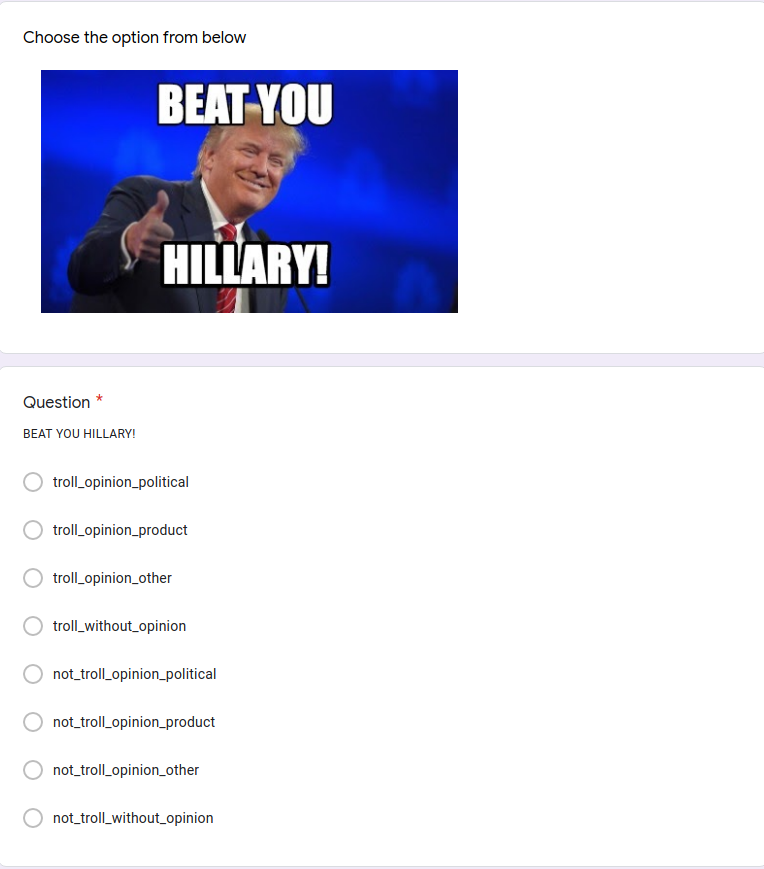} 
  \caption{An example from google form that provides the meme and option to choose from}
    \label{fig:Example_pred} 
\end{figure*}

In pilot studies, we used 24 volunteers to annotate which belonged to a different background -- in terms of gender, education, and nationality as mentioned in Table \ref{tab:annotators}. The similarity between annotations and gold labels (annotated by author) has been measured with the Jaccard similarity score.

Below are the findings from the pilot studies:
\begin{enumerate}
\renewcommand{\theenumi}{\Roman{enumi}}
\renewcommand{\labelenumi}{\theenumi}
    \item The choice or label that has been given by each annotator is subjective. This is the root cause of disagreement which understandable as memes are vague and directed towards a particular individual or group that shares similar ideas. 
    \item The initial choices for the annotations were non\_troll, troll\_without\_opinion, troll\_opinion\_x\_positive, troll\_opinion\_x\_negative and troll\_opinion\_x\_neutral where x could be product, political or personal. Later, we changed classes to ones mentioned in section \ref{Troll} because word troll is generally negative and it cannot be associated with anything positive or neutral.
\end{enumerate}

For the final work, we chose two male and two females with top Jaccard similarity score\footnote{\href{https://scikit-learn.org/stable/modules/generated/sklearn.metrics.jaccard_score.html} Jaccard Similarity}. 
Initially, 10 google forms were provided to each annotator, and later, based on their delay in response and willingness, we provided the next batch of google forms for annotation. This process was effective in terms of coverage (number of samples annotated by at least one annotator), and to facilitate the majority vote we maintained three annotations per sample. As a result, we covered all the data but resulted in empty values for one annotation per sample. Hence we needed an agreement which could accommodate empty data. In this case, Krippendorff's alpha turned out to be a reliable inter-annotator agreement scheme.

\subsection{Inter-annotator Agreement}
As discussed previously, we needed an agreement which can accommodate all four annotators along with incomplete data. In this case, Krippendorff's alpha \citewopar{krippendorff2011computing} turned out to be useful as it deals with this issue. Krippendorff's alpha calculates the reliability coefficient that measures the agreement amongst annotators by distinguishing between the incomplete units (annotations) or assigning computable values to them. It is a ratio of observed disagreement $D_o$ and expected disagreement $D_e$ (see Equation 1), that ranges from 0 to 1, where 0 means perfect disagreement, and 1 means perfect agreement. 
\begin{equation}
    \alpha = 1 - \frac{D_o}{D_e}
\end{equation}
 \begin{equation}
    D_o = \frac{1}{n}\sum_{c}\sum_{k}o_{ck\;metric}\;\delta^2_{ck}
\end{equation}
\begin{equation}
D_e = \frac{1}{n(n-1)} \sum_{c}\sum_{k}n_c \; .\;n_{k\;metric}\,\delta^2_{ck}
\end{equation}
Here $o_{ck}\;n_c\;n_k\;$ and $n$ indicates the frequencies of values in the coincidence matrices and $metric$ indicates a metric or level of measurement such as nominal, ordinal, interval, ratio and others. Krippendorff's alpha applies to all these metrics. In our case, we refered nominal and interval metric to calculate inter-annotator agreement.

We were able to achieve significant agreement amongst annotators (nominal metric: 0.699; interval metric: 0.775). The final label for each class has been decided based on the majority vote. In case of conflict, expert's (authors) vote considered as final.

\subsection{Difficult Examples}
Memes are inherently obscure and meant for the targeted group. We observed this issue while going through annotated samples. Some samples had interesting annotation as mentioned below: 

\begin{enumerate}
\renewcommand{\theenumi}{\Roman{enumi}}
\renewcommand{\labelenumi}{\theenumi}
    \item An example circled \circled{G} from Figure \ref{fig:Example_not_troll} was annotated as Troll\_opinion\_product by one annotator, as it is talking about Minecraft (product) negatively. 
    \item An example circled \circled{D} from Figure \ref{fig:Example_product} was annotated as Not\_troll\_without\_opinion as it is motivational if one ignores that it is associated with a company (Apple\footnote{Apple Inc. is an American multinational technology company headquartered in Cupertino, California, that designs, develops and sells consumer electronics, computer software, and online services.}).
    \item An example circled \circled{H} from Figure \ref{fig:Example_not_troll} was annotated as Not\_troll\_opinion as it is portraying the original poster (one who posted) positively
\end{enumerate}

The annotation guidelines are helpful, but it cannot be a formula or solution that fits for all, and also there is no one correct label for each meme. Hence, this task could become a multi-class, multi-label. We leave this multi-class, multi-label approach to future work.

% \begin{table}[!htb]  
% \begin{center} 
% % \renewcommand{\tabcolsep}{1.5mm}
% \captionsetup{font=normalsize}
% \normalsize
% \begin{tabular}{|l|l|l|l|}
% \hline
% & & Initial/Pilot Study & Final Study \\
% \hline
% Gender & Male & 14 & 2 \\
% & Female & 9 & 2 \\
% & Non-binary & 1 & - \\
% \hline
% Education & Undergraduate & 4 & - \\
% & Graduate & 8 & 2 \\
% & Postgraduate & 12 & 2 \\
% \hline
% Total & - & 24 & 4 \\
% \hline
% \end{tabular}
% \caption{Initial Study } 
% \label{tab:annotators}
% \end{center} 
% \end{table}

\subsection{Annotators}
We selected annotators from the pool of volunteers based on the Jaccard similarity index. After finalizing four annotators, we did the second round of annotation (final study) and selected the label based on the majority vote. In a case of conflict, we went to the expert for resolution. 

In our pilot study, we saw volunteers from different demographics in terms of nationality (UK, Germany, Australia, USA, India), gender (male, female, non-binary) and education (undergraduate, graduate, postgraduate). Table \ref{tab:annotators} shows the distribution of these demographics between the pilot study and final work. Distribution of demographics in Table \ref{tab:annotators} shows the dominance of the age group of 25 to 34 in both phases (pilot study and final work). This involvement could be attributed to the rise of meme culture during the era of Millenials, as they are the ones who started this culture. As a result, we see less involvement of other age groups -- most of them cant connect with meme culture.
Due to resource constraints, we cannot include people from all demographics to understand each meme. Hence we relied on randomly allocated volunteers. This randomization, with the Jaccard similarity index, helped us to achieve a significant agreement amongst annotators.

\begin{table}[!htb]  
\begin{center} 
\renewcommand{\tabcolsep}{1.7mm}
% \captionsetup{font=normalsize}
\normalsize
\begin{tabular}{|l|l|r|r|}
\hline
\multicolumn{2}{|c|}{Phase} & Pilot & Final\\
\hline
Gender & Male & 14 & 2\\
& Female & 9 & 2\\
& Non-binary & 1 & -\\
\hline
Education & Undergraduate & 4 & -\\
& graduate & 8 & 2\\
& Postgraduate & 12 & 2\\
\hline
Age group    & 18 to 24 & 4 & 1\\
& 25 to 34 & 15 & 3\\
& 35 to 44 & 3 & -\\
& 45 to 54 & 1 & -\\
& 55 to 64 & 1 & -\\
\hline
\multicolumn{2}{|c|}{ Total } & 24 & 4\\
\hline
\end{tabular}
\caption{Distribution of demographics (gender, education and age-group) for annotators in pilot study and final work} 
\label{tab:annotators} 
\end{center} 
\end{table}

\section{Data Statistics}

\begin{table}[t]
\begin{center} 
\begin{tabular}{|l|r|r|r|r|}
\hline
Parameters (per sample) & Mean & Standard deviation & Minimum value & Maximum value\\
\hline
% Number of samples & 6292 \\
Word count & 16.55 & 14.61 & 1.00 & 306.00\\
% Avg character count per tweet & 80.83 \\
Stopword count & \textbf{2.73} & 5.15 & 0.00 & 100.00\\
Capital word count & 7.15 & 7.60 & 0.00 & 76.00\\
Numerical value count & 0.20 & 0.96 & 0.00 & 56.00\\
Character count & 91.49 & 76.69 & 2.00 & 1600.00\\
% Avg numeric values per tweet & 0.07 \\ 
\hline
\end{tabular} 
\caption{Data statistics for the text associated with the meme from the TrollWithOpinion dataset} 
\label{tab:data_stat_text} 
\end{center} 
\end{table}

\begin{table}[t]
\begin{center} 
\begin{tabular}{|l|r|}
\hline
Class & Distribution\\
\hline
troll\_opinion\_political & 338 \\
not\_troll\_without\_opinion & 847 \\
not\_troll\_opinion\_political & 253 \\
troll\_opinion\_other & 1,103 \\
not\_troll\_opinion\_other & 1,618 \\
not\_troll\_opinion\_product & 779 \\
troll\_without\_opinion & 3,782 \\
troll\_opinion\_product & 161 \\
\hline
Total & 8,881 \\
% Avg numeric values per tweet & 0.07 \\ 
\hline
\end{tabular} 
\caption{Data distribution for the classes in the TrollWithOpinion dataset} 
\label{tab:data_distribution} 
\end{center} 
\end{table}

\begin{figure}[tbh]
  \centering
  \begin{minipage}[b]{1.0\textwidth}
    \includegraphics[width=\textwidth, height=7cm]{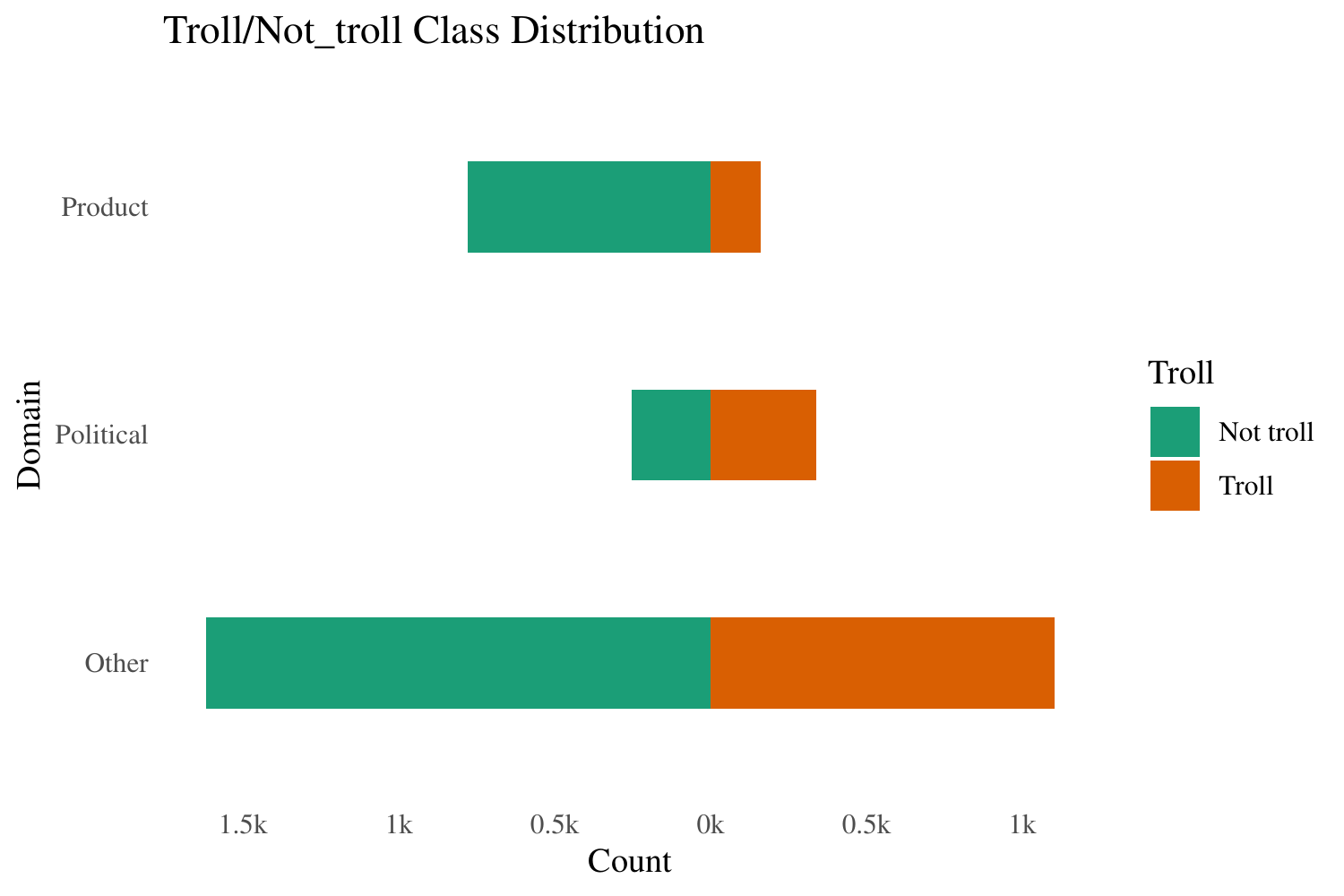}
    \caption{Domain-wise distribution of troll and not troll classes in the TrollWithOpinion dataset.}
    \label{fig:data_distribution_1} 
  \end{minipage}
  \hfill
  \begin{minipage}[b]{1.0\textwidth}
    \includegraphics[width=\textwidth, height=7cm]{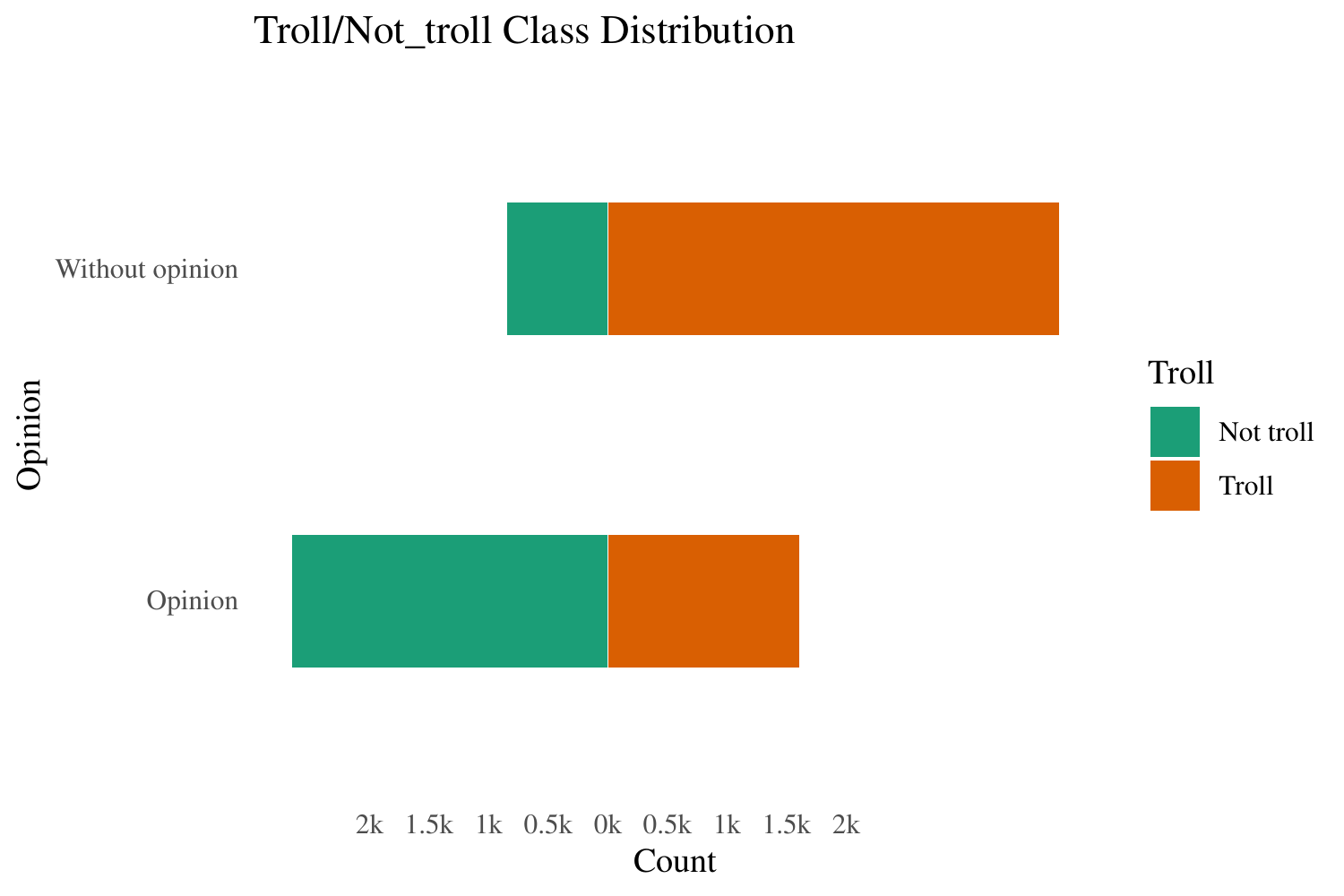}
    \caption{Opinion-wise distribution of troll and not troll classes the TrollWithOpinion dataset.}
    \label{fig:data_distribution_2} 
  \end{minipage}
\end{figure}

Table \ref{tab:data_stat_text} shows an overview of the data statistics (Mean, Standard deviation, Minimum value, Maximum value) for parameters -- word count, stop word count, capital word count, numerical value count and character count per sample -- of the text associated with memes. On average, out of 16.55 words, 2.73 are stopwords. However, numerical values account for only 0.20 out of 16.55 words. Hence, we can say that stopwords \footnote{\href{https://www.nltk.org/book/ch02.html}{nltk English stopwords}} might hold the most valuable information, while numerical values the least. But this is based on a text-level analysis, while on a corpus-level (text from all the memes), stopwords are of negligible interest because they occur with greater frequency; thus keeping them after text processing may not be useful. Furthermore, we can see that there is a high standard deviation for each parameter. In the case of ``Word count'', the standard deviation of 14.61 suggests that the length of the text varies by the word count of 14 from the mean value i.e. 16.55 on an average. Also, the text length could vary from minimum 1 to maximum 306. The maximum word count is helpful in limiting the length of the text. In our case, we limited the length of the text to 300 based on this statistic.

Table \ref{tab:data_distribution} illustrates the distribution of classes in the annotated corpus. As we are annotating existing dataset, the distribution of the classes shows an imbalance. If memes were collected using keyword searches as per classes, then a more balanced class distribution might have been achieved. However, such a balanced distribution might not represent the data in the wild (internet). While developing the memotion dataset, the authors collected data from 52 unique and globally popular categories. This bias will help us to understand the distribution of proposed classes (from Table \ref{tab:data_distribution}) in popular (majorly talked) opinions on the internet.

It is worth noting that the troll\_without\_opinion class dominates the class distribution presented in Table \ref{tab:data_distribution}. The marginal gap between this and the second most common class, not\_troll\_opinion\_other, is significant. The prevalence of the troll\_without\_opinion might be due to the difficulty in finding opinion manipulation, as not all meme intentions are explicit. We plan to mitigate against this in future by collecting more annotations, which can potentially turn this task into a multi-label, multi-class classification problem.

Figure \ref{fig:data_distribution_1} illustrates the domain-wise (product, political or other) distribution of troll and not-troll classes irrespective of the presence of opinion manipulation. As shown, the number of troll classes -- troll\_opinion\_X (X could be political, product or other) except troll\_without\_opinion -- is lower compared to that of not-troll classes -- not\_troll\_opinion\_X except not\_troll\_without\_opinion. This disparity could be due to the fact that there are fewer troll memes in popular opinions generally. Figure \ref{fig:data_distribution_1} also shows that the highest number of troll memes are present in other domains irrespective of the presence of opinion manipulation. It is not a surprise that a higher number of troll memes than not-troll memes are present in the political domain. Hence, a meme is more likely to be a troll in popular opinions if it is political. In the product domain, a higher number of not-troll than troll memes are present. 

Figure \ref{fig:data_distribution_2} shows the distribution of opinion-wise (opinion and without opinion) distribution of troll and not-troll classes irrespective of domain. This graph shows that a higher number of troll memes are present in without-opinion class when compared with opinion class. As without-opinion classes do not belong to any of the domains, it is absent from Figure \ref{fig:data_distribution_1}; however, these classes are included in the opinion-wise distribution of troll and not-troll classes. 
We can summarise that there tends to be a higher number of not-troll memes in the opinion class, while there is a higher number of troll memes in the without-opinion class.

\section{Experiments}
This section describes all the baseline experiments performed on the newly developed TrollWithOpinion dataset. This section describes six baselines (laid out by \citewopar{kiela2019supervised}) that we used to model our TrollMemeOpinion dataset along with five machine learning baselines. The dataset was split into train-test-validation with 80\%, 10\% and 10\% split ratio. All the baselines were evaluated on the held-out test set with 896 samples.

\subsection{Traditional Machine Learning baselines}

For all the traditional machine learning (ML) baselines, we used state-of-the-art contextual BERT \citewopar{devlin2018bert} embeddings. These BERT embeddings are derived from the last hidden state of DistilBERT \citewopar{sanh2019distilbert}. As a result, a 768-dimensional vector representation of a text has been achieved. Later, this text vector has been used as an input to each traditional machine learning baseline. To ensure the optimal performance by each of these ML baselines, we incorporated a grid search to find suitable hyperparameters with the help of four-fold cross-validation technique \citewopar{shao1993linear}. L2 regularization has been used to avoid overfitting.

The general de-facto method of applying ML to the text starts by the vectorization of the text. Instead of using a tf-idf vector representation of the text, we used the last hidden layer representation from DistillBert with 768 hidden units i.e. 768 dimensional (768 d) features. This contextual representation of the text acts like 768 independent features used to predict the targeted independent variable i.e. if the text is offensive (OFF) or not (NOT). Below, we explained each ML baselines. Please note that the 768 d features were derived from Distillbert were not fine tuned, instead we froze them.

\paragraph{Logistic regression} (LR) is a basic building block to a neural network. By keeping this in mind, we decided to leverage its simple architecture to classify a given meme based on just the text (associated with meme). Our LR assumes a linear relation between text and label, with this assumption, it predicts the best line of fit for given text features (768-dimensional vector from BERT). It takes in these text features and calculates the sigmoid function which later has been translated into 1 (OFF) or 0 (NOT) based on the threshold value i.e. 0.5. As we assume that these text features are linearly dependent, we do not use any kind of nonlinearity such as  ReLU or GeLU. We optimized the algorithm based on the grid search over the inverse regularization strength.

\paragraph{Gaussian Naive Bayes} (GNB) \citewopar{zhang2005exploring} is a probabilistic ML baseline that naively assumes that each feature from 768 features is conditionally independent of others. The Naive Bayes variation used by us (GNB) assumes the underlying likelihood function to be a Gaussian distribution. Posteriori probability i.e. probability of independent label given the 768d features is calculated by multiplying the Gaussian likelihood function. We used the default parameter setting from sklearn library with the portion of largest variance of all features set to 1e\-9.

\paragraph{Support vector machine} (SVM) \citewopar{boser1992training} is a discriminative classification that separates data points using a hyperplane; the objective of the classifier is to keep these hyperplanes separated from each other with large margins. But the traditional version of SVM does not scale well with a large dataset, hence we are using linear SVM \citewopar{fan2008liblinear} with flexible penalties and loss functions. It uses a linear kernel, and parameters are tuned based on the hinge loss function. We trained a linear SVM with the random state of 0 (to reproduce the results) with the tolerance for stopping criteria of 1e\-5.

\paragraph{Random forest} (RF) \citewopar{breiman2001random} is an ensemble of decision tree classifier where the data samples are distributed amongst the suitable decision trees (DT) \citewopar{breiman1984classification} and their outcomes are averaged to improve the evaluation metric scores. Each decision tree predicts the target label by learning if-then-else decision rules. We chose RF to leverage the scaled simplicity of the DT in the form of ensemble architecture. We trained RF with the grid search over the 50 to 200 range of number of trees along with both Gini impurity as well as entropy as a criteria to measure the quality of the split.

\paragraph{K-nearest neighbour} (KNN) \citewopar{altman1992introduction} classifier is a non-generalizing algorithm that assigns a class label based on the majority vote from the K-nearest neighbours. We used KNN as a baseline to classify the text associated with the meme into the required classes. We did a grid search for the value of K in the [2,4,6,8]. In this baselines, neighbours are derived based on the euclidean distance between them.

\subsection{Deep Learning baselines}

% \paragraph{Bidirectional text LSTM} (with GloVe \citewopar{pennington2014glove}) initialized with ResNet image embeddings (BiLSTM + ResNet) architecture \footnote{https://github.com/04mayukh/Memebusters-at-SemEval-2020-Task-8-Memotion-Analysis} takes feature maps from ResNet as an initial state of BiLSTM, later adding dot-product attention to the architecture. 

% \paragraph{Parallel-Channel Model} \citewopar{yuan2020ynu} consists of two channels, one each for text and image. This architecture proposes three models in the text channel, and two in the image channel. Later, the outcomes of these models are forwarded to a soft voting mechanism which decides the outcome.

% \paragraph{BERT-DenseNet} \citewopar{guonuaa}: Combines BERT representation of text and image representation from DenseNet201 which connects each layer to every other layer in a feed-forward fashion.

% \paragraph{A Robustly Optimized BERT Pretraining Approach (RoBERTa) + ResNet} \citewopar{gupta2020dsc}is an early fusion technique that concatenates the textual and visual feature vectors extracted by RoBERTa and ResNet respectively to produce a joint multimodal representation before passing it through a softmax classifier.

\paragraph{Bag of words (BOW)}: This unimodal baseline used text representation in the form of 300-dimensional pre-trained GloVe \citewopar{pennington2014glove} (common crawl) word embeddings, which were fed to a classification layer. These GloVe embeddings were fine-tuned on word-word co-occurrence statistics from a text corpus curated from our dataset. We chose this unimodal text baseline to analyze the classification performance with the use of non-contextual word embedding.

\paragraph{Text-only BERT (BERT)}: In this unimodal baseline, only text modality was used to train a pre-trained base uncased version of BERT to get the BERT representation of given text. This BERT representation was provided to a fully connected network which formed a classification layer. 
This model is uncased as it does not make a difference between uppercase and lowercase words. Unlike sequential models, Recurrent Neural Network (RNN), this transformer-based architecture does not rely on the sequence of the words. Instead, it was trained in Masked Language Modelling (MLM) fashion, which randomly masks 15\% of the words from the input text, and later predict these masked words. 
This way, the inner representation (BERT features) of the text were learnt which later were finetuned on the text corpus curated from our TrollsWithOpinion dataset. Unlike GloVe, BERT embeddings derived from BERT features are contextual. 

\paragraph{Image-only (Img)}: This unimodal baseline made use of image representation derived from ResNet-152 \citewopar{he2016deep} pre-trained on imagenet. Later, this image representation with a 2048-dimensional feature vector was fed to a classification layer. ResNet-152 is capable of learning feature maps that are not prone to the vanishing gradient descent problem, the problem that is generally seen in the plain deep convolution networks (without residuals). The ``shortcut connections" or ``skip connections" proposed by this architecture does the identity mapping without adding extra parameters or complexity to the neural network. 

\paragraph{Concat BOW + Img (ConcatBow)}: In this multimodal baseline, both text and image representations were used in the form concatenated feature vectors which were derived from BOW (text) and Img (image) baselines. This concatenation of features resulted in a 2048+300-dimensional feature vector, which was fed to a classification layer. This architecture follows an early fusion technique wherein text featured from GloVe and image features from ResNet-152 were concatenated and trained jointly by feeding to a sigmoid layer.

\paragraph{Concat BERT + Img (ConcatBert)}: In this multimodal baseline, text representation from BERT and image representation from Img baselines were concatenated to form a 2048+768-dimensional feature vector. Later, this feature vector was fed to a classification layer with a sigmoid activation function (as with all baselines). The difference between ConcatBow and ConcatBert comes from textual features while the earlier one uses GloVe (non-contextual word embedding) and the later one uses BERT (contextual word embeddings). 

\paragraph{Multimodal Bitransformer (MMBT)}: This multimodal baseline used both image and text features derived from Img and BERT baselines which were later combined using bidirectional transformer architecture similar to BERT. The architecture takes in the contextual word embeddings in the form of text features derived from BERT baselines which were added to the segment, positional embeddings. On the other hand, image features derived from ResNet were mapped to the semantic space of the contextual embeddings to form image tokens which were again added with the segment and positional embeddings before feeding to Bi-directional transformers. This architecture emphasizes that the textual features carry more meaningful information than the image features, which is the reason behind mapping image features in the semantic space of word features.

\subsection{Hyperparameter Settings}

We trained each baseline for 20 epochs with patience set at 15 epochs. The initial learning rate was the same -- i.e. 1e\-07 for each baseline -- but later it was changed dynamically based on the validation accuracy of the baseline. If the learning stagnated (no improvement in validation accuracy), then the learning rate was dropped by a factor of 0.5 of the previous learning rate. Also, in the case of multimodal baselines, we froze all the hidden layers associated with the text and image models for the first five and ten epochs respectively.

\begin{figure*}[tp]
  \centering 
  \includegraphics[width=1\linewidth,clip]{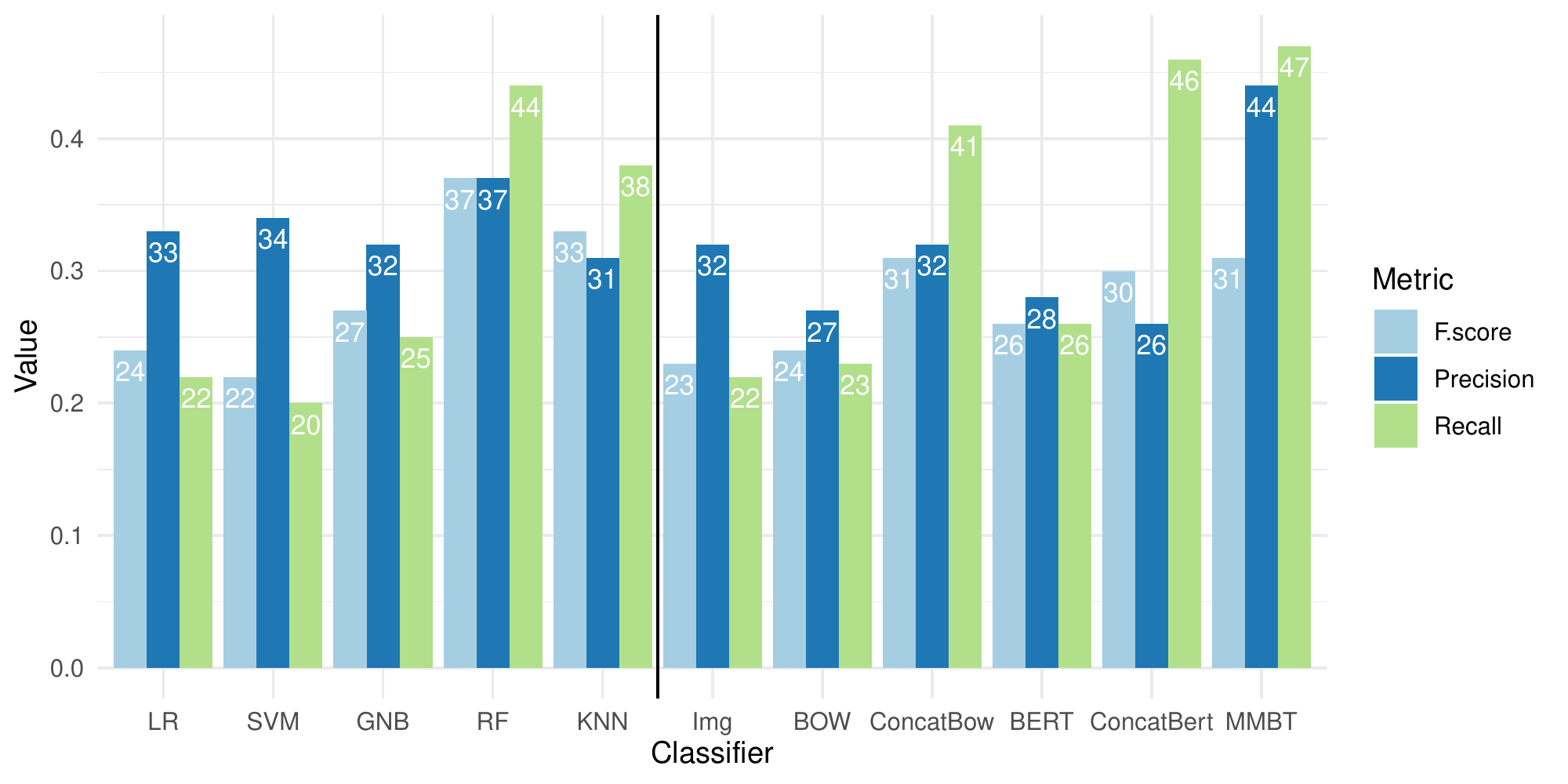} 
  \caption{A bargraph for comparison of weighted-average precision, recall and F-score across ML (left hand side) and DL (right hand side) baselines.}
    \label{fig:weighted} 
\end{figure*}

\begin{table*}[tp]
\centering
\begin{tabular} {l|r|r|r|r|r|r|}%{l|lllllllll}
\hline
 	& \multicolumn{6}{c}{\textbf{Precision}}\\
 	\hline	
 	 & \multicolumn{1}{c|}{\textbf{LR}} & \multicolumn{1}{c|}{\textbf{SVM}} & \multicolumn{1}{c|}{\textbf{GNB}} & \multicolumn{1}{c|}{\textbf{RF}} & \multicolumn{1}{c|}{\textbf{KNN}} & \multicolumn{1}{c|}{\textbf{Count}}\\
 	\hline
	\texttt{Troll\_opinion\_political} & 0.09 & 0.09 & 0.10 & 0.12 & 0.00 & 26 \\
	\texttt{Not\_troll\_opinion\_political} & 0.09 & 0.09 & 0.07 & 0.00 & 0.00 & 24 \\
	\texttt{Troll\_opinion\_product} & 0.04 & 0.02 & 0.02 & 0.00 & 0.00 & 15 \\
	\texttt{Not\_troll\_opinion\_product} & 0.15 & 0.15 & 0.10 & 0.28 & 0.24 & 61 \\
	\texttt{Troll\_opinion\_other} & 0.18 & 0.24 & 0.19 & 0.27 & 0.15 & 110 \\
	\texttt{Not\_troll\_opinion\_other} & 0.32 & 0.30 & 0.26 & 0.39 & 0.26 & 167 \\
	\texttt{Troll\_without\_opinion} & 0.49 & 0.48 & 0.48 & 0.49 & 0.46 & 410 \\
	\texttt{Not\_troll\_without\_opinion} & 0.17 & 0.16 & 0.16 & 0.13 & 0.23 & 83 \\
	\hline
% 	Macro avg & 0.19 & 0.19 & 0.17 & 0.21 & 0.17 & 896 \\
	Weighted avg & 0.33 & 0.34 & 0.32 & 0.37 & 0.31 & 896 \\
	\hline
	& \multicolumn{6}{c}{\textbf{Recall}}\\
	\hline
	& \multicolumn{1}{c|}{\textbf{LR}} & \multicolumn{1}{c|}{\textbf{SVM}} & \multicolumn{1}{c|}{\textbf{GNB}} & \multicolumn{1}{c|}{\textbf{RF}} & \multicolumn{1}{c|}{\textbf{KNN}} & \multicolumn{1}{c|}{\textbf{Count}}\\
 	\hline
	\texttt{Troll\_opinion\_political} & 0.27 & 0.19 & 0.19 & 0.08 & 0.00 & 26 \\
	\texttt{Not\_troll\_opinion\_political} & 0.21 & 0.33 & 0.17 & 0.00 & 0.00 & 24 \\
	\texttt{Troll\_opinion\_product} & 0.13 & 0.27 & 0.07 & 0.00 & 0.00 & 15 \\
	\texttt{Not\_troll\_opinion\_product} & 0.33 & 0.26 & 0.16 & 0.13 & 0.13 & 61 \\
	\texttt{Troll\_opinion\_other} & 0.25 & 0.11 & 0.25 & 0.06 & 0.08 & 110 \\
	\texttt{Not\_troll\_opinion\_other} & 0.26 & 0.14 & 0.22 & 0.25 & 0.24 & 167 \\
	\texttt{Troll\_without\_opinion} & 0.18 & 0.20 & 0.31 & 0.80 & 0.66 & 410 \\
	\texttt{Not\_troll\_without\_opinion} & 0.28 & 0.31 & 0.18 & 0.07 & 0.11 & 83 \\
	\hline
% 	Macro avg & 0.24 & 0.23 & 0.19 & 0.17 & 0.15 & 896 \\
	Weighted avg & 0.22 & 0.20 & 0.25 & 0.44 & 0.38 & 896 \\
	\hline
	& \multicolumn{6}{c}{\textbf{F-score}}\\
	\hline
	 	 & \multicolumn{1}{c|}{\textbf{LR}} & \multicolumn{1}{c|}{\textbf{SVM}} & \multicolumn{1}{c|}{\textbf{GNB}} & \multicolumn{1}{c|}{\textbf{RF}} & \multicolumn{1}{c|}{\textbf{KNN}} & \multicolumn{1}{c|}{\textbf{Count}}\\
 	\hline
	\texttt{Troll\_opinion\_political} & 0.14 & 0.13 & 0.14 & 0.09 & 0.00 & 26 \\
	\texttt{Not\_troll\_opinion\_political} & 0.13 & 0.15 & 0.10 & 0.00 & 0.00 & 24 \\
	\texttt{Troll\_opinion\_product} & 0.06 & 0.04 & 0.03 & 0.00 & 0.00 & 15 \\
	\texttt{Not\_troll\_opinion\_product} & 0.20 & 0.19 & 0.12 & 0.18 & 0.17 & 61 \\
	\texttt{Troll\_opinion\_other} & 0.21 & 0.15 & 0.22 & 0.10 & 0.10 & 110 \\
	\texttt{Not\_troll\_opinion\_other} & 0.28 & 0.19 & 0.24 & 0.31 & 0.25 & 167 \\
	\texttt{Troll\_without\_opinion} & 0.26 & 0.28 & 0.37 & 0.61 & 0.54 & 410 \\
	\texttt{Not\_troll\_without\_opinion} & 0.21 & 0.21 & 0.17 & 0.09 & 0.15 & 83 \\
	\hline
% 	Macro avg & 0.19 & 0.17 & 0.17 & 0.17 & 0.15 & 896 \\
	Weighted avg & 0.24 & 0.22 & 0.27 & 0.37 & 0.33 & 896 \\
	\hline
\end{tabular}
\caption{Precision, Recall, and F-score for Machine Learning baselines trained on the TrollWithOpinion dataset.}
\label{tab:result_ML}
\end{table*}

\begin{table*}[tp]
\centering
\begin{tabular} {l|r|r|r}%{l|lllllllll}
\hline
 	& \multicolumn{3}{c}{\textbf{Precision}}\\
 	\hline	
 	 & \multicolumn{1}{c|}{\textbf{Img}} & \multicolumn{1}{c}{\textbf{BOW}} & \multicolumn{1}{c|}{\textbf{ConcatBow}}\\
 	\hline
	\texttt{Troll\_opinion\_political} & 0.01 & 0.00 & 0.00 \\
	\texttt{Not\_troll\_opinion\_political} & 0.04 & 0.03 & 0.00 \\
	\texttt{Troll\_opinion\_product} & 0.00 & 0.00 & 0.00 \\
	\texttt{Not\_troll\_opinion\_product} & 0.00 & 0.14 & 0.67 \\
	\texttt{Troll\_opinion\_other} & 0.11 & 0.07 & 0.00 \\
	\texttt{Not\_troll\_opinion\_other} & 0.38 & 0.20 & 0.19 \\
	\texttt{Troll\_without\_opinion} & 0.48 & 0.45 & 0.45 \\
	\texttt{Not\_troll\_without\_opinion} & 0.10 & 0.09 & 0.29\\
	\hline
% 	Macro avg & 0.14 & 0.12 & 0.20 & 0.12 & 0.09 & 0.25\\
	Weighted avg & 0.32 & 0.27 & 0.32\\
	& \multicolumn{3}{c}{\textbf{Recall}}\\
	\hline
	& \multicolumn{1}{c|}{\textbf{Img}} & \multicolumn{1}{c}{\textbf{BOW}} & \multicolumn{1}{c|}{\textbf{ConcatBow}} \\
 	\hline
	\texttt{Troll\_opinion\_political} & 0.04 & 0.00 & 0.00\\
	\texttt{Not\_troll\_opinion\_political} & 0.38 & 0.33 & 0.00\\
	\texttt{Troll\_opinion\_product} & 0.00 & 0.00 & 0.00\\
	\texttt{Not\_troll\_opinion\_product} & 0.00 & 0.05 & 0.07\\
	\texttt{Troll\_opinion\_other} & 0.11 & 0.05 & 0.00\\
	\texttt{Not\_troll\_opinion\_other} & 0.03 & 0.18 & 0.10\\
	\texttt{Troll\_without\_opinion} & 0.40 & 0.38 & 0.81\\
	\texttt{Not\_troll\_without\_opinion} & 0.07 & 0.04 & 0.11\\
	\hline
% 	Macro avg & 0.13 & 0.13 & 0.14 & 0.11 & 0.13 & 0.13\\
	Weighted avg & 0.22 & 0.23 & 0.41\\
	\hline
	& \multicolumn{3}{c}{\textbf{F-Score}}\\
	\hline
	& \multicolumn{1}{c}{\textbf{BERT}} & \multicolumn{1}{c|}{\textbf{ConcatBert}} & \multicolumn{1}{c|}{\textbf{MMBT}}\\
 	\hline
	\texttt{Troll\_opinion\_political} & 0.00 & 0.00 & 0.00\\
	\texttt{Not\_troll\_opinion\_political} & 0.00 & 0.00 & 0.00\\
	\texttt{Troll\_opinion\_product} & 0.03 & 0.00 & 0.00\\
	\texttt{Not\_troll\_opinion\_product} & 0.08 & 0.00 & 0.00\\
	\texttt{Troll\_opinion\_other} & 0.02 & 0.00 & 0.00\\
	\texttt{Not\_troll\_opinion\_other} & 0.20 & 0.05 & 0.05\\
	\texttt{Troll\_without\_opinion} & 0.46 & 0.63 & 0.63\\
	\texttt{Not\_troll\_without\_opinion} & 0.00 & 0.00 & 0.11\\
	\hline
% 	Macro avg & 0.10 & 0.10 & 0.12 & 0.10 & 0.09 & 0.10\\
	Weighted avg & 0.26 & 0.30 & 0.31\\
	\hline
 	& \multicolumn{3}{c}{\textbf{Precision}}\\
 	\hline	
 	 & \multicolumn{1}{c}{\textbf{BERT}} & \multicolumn{1}{c|}{\textbf{ConcatBert}} & \multicolumn{1}{c|}{\textbf{MMBT}} \\
 	\hline
	\texttt{Troll\_opinion\_political} & 0.00 & 0.00 & 0.00\\
	\texttt{Not\_troll\_opinion\_political} & 0.00 & 0.00 & 0.00\\
	\texttt{Troll\_opinion\_product} & 0.01 & 0.00 & 0.00\\
	\texttt{Not\_troll\_opinion\_product} & 0.09 & 0.00 & 0.00\\
	\texttt{Troll\_opinion\_other} & 0.20 & 0.00 & 0.00\\
	\texttt{Not\_troll\_opinion\_other} & 0.17 & 0.26 & 1.00\\
	\texttt{Troll\_without\_opinion} & 0.48 & 0.46 & 0.46\\
	\texttt{Not\_troll\_without\_opinion} & 0.00 & 0.00 & 0.5\\
	\hline
% 	Macro avg & 0.14 & 0.12 & 0.20 & 0.12 & 0.09 & 0.25\\
	Weighted avg 0.28 & 0.26 & 0.44\\
	\hline
		& \multicolumn{3}{c}{\textbf{Recall}}\\
	\hline
	& \multicolumn{1}{c}{\textbf{BERT}} & \multicolumn{1}{c|}{\textbf{ConcatBert}} & \multicolumn{1}{c|}{\textbf{MMBT}}\\
 	\hline
	\texttt{Troll\_opinion\_political} & 0.00 & 0.00 & 0.00\\
	\texttt{Not\_troll\_opinion\_political} & 0.00 & 0.00 & 0.00\\
	\texttt{Troll\_opinion\_product} & 0.13 & 0.00 & 0.00\\
	\texttt{Not\_troll\_opinion\_product} & 0.07 & 0.00 & 0.00\\
	\texttt{Troll\_opinion\_other} & 0.01 & 0.00 & 0.00\\
	\texttt{Not\_troll\_opinion\_other} & 0.25 & 0.03 & 0.02\\
	\texttt{Troll\_without\_opinion} & 0.44 & 0.98 & 1.00\\
	\texttt{Not\_troll\_without\_opinion} & 0.00 & 0.00 & 0.06\\
	\hline
% 	Macro avg & 0.13 & 0.13 & 0.14 & 0.11 & 0.13 & 0.13\\
	Weighted avg & 0.26 & 0.46 & 0.47\\
	\hline
		& \multicolumn{3}{c}{\textbf{F-Score}}\\
	\hline
	& \multicolumn{1}{c}{\textbf{BERT}} & \multicolumn{1}{c|}{\textbf{ConcatBert}} & \multicolumn{1}{c|}{\textbf{MMBT}}\\
 	\hline
	\texttt{Troll\_opinion\_political} & 0.00 & 0.00 & 0.00\\
	\texttt{Not\_troll\_opinion\_political} & 0.00 & 0.00 & 0.00\\
	\texttt{Troll\_opinion\_product} & 0.03 & 0.00 & 0.00\\
	\texttt{Not\_troll\_opinion\_product} & 0.08 & 0.00 & 0.00\\
	\texttt{Troll\_opinion\_other} & 0.02 & 0.00 & 0.00\\
	\texttt{Not\_troll\_opinion\_other} & 0.20 & 0.05 & 0.05\\
	\texttt{Troll\_without\_opinion} & 0.46 & 0.63 & 0.63\\
	\texttt{Not\_troll\_without\_opinion} & 0.00 & 0.00 & 0.11\\
	\hline
% 	Macro avg & 0.10 & 0.10 & 0.12 & 0.10 & 0.09 & 0.10\\
	Weighted avg & 0.26 & 0.30 & 0.31\\
	\hline
\end{tabular}
\caption{Precision, Recall, and F-score for Deep Learning baselines trained on the TrollWithOpinion dataset.}
\label{tab:result_DL}
\end{table*}

\section{Results and Discussion}

Table \ref{tab:result_ML} and Table \ref{tab:result_DL} shows the detailed class-wise evaluation metric (precision, recall, F-score) for each machine learning (ML) and deep learning (DL) baseline. We chose the weighted-average evaluation metric over macro-average to address the class imbalance problem.

Figure \ref{fig:weighted} is a bar graph that shows the comparison of weighted-average precision, recall and F-score amongst ML and DL baselines. Here, the x-axis represents 11 baselines (five ML and six DL) separated by a vertical line, and the y-axis represents the value of each evaluation metric. Overall, we can see that the ConcatBow and its unimodal counterpart (i.e. BOW) shows the top two highest weighted precision scores, 0.47 and 0.41, respectively. In this particular case, the inclusion of image features aided in the improvements of weighted-average precision. On the contrary, the weighted-average precision declines on the inclusion of image features in the case of BERT (precision - 0.27) and ConcatBert (precision - 0.22). This contradiction could be attributed to the complexity of the model. BERT being more complex (contextual word representation) than the BOW (non-contextual word representation), and hence needs more data to train. Furthermore, the Img baseline falls on the third position with 0.38 weighted-average precision, shows that only image feature from the meme could also be used to identify the classes. But the poor recall (0.22) of the Img baseline shows the prevalence of false negatives which is undesirable.

Table \ref{tab:result_ML} and Table \ref{tab:result_DL} shows that the evaluation metric for Troll\_without\_opinion is higher compared to any other classes irrespective of ML baseline. Furthermore, all the DL baselines show a similar trend. This could be attributed to the imbalance in the dataset as Troll\_without\_opinion contributes to 42 \% of the total data. On the hand, all ML and DL baseline perform poorly on Troll\_opinion\_product which could be attributed to the fact there are the least number (1 \%) of samples in the dataset for this class. As per the precision, recall and F-score, the Not\_troll\_opinion\_product is more likely to get identified by ML baselines more than any of the DL baselines. Overall, we either need more data or an approach prone to the data imbalance.

\section{Conclusion}

Our work introduced a meme dataset with hierarchical annotation scheme that studies the effect of troll meme in the form of opinion manipulation. We introduced eight meme classes and enhanced the existing dataset to form the TrollsWithOpinion dataset. 
We found out that textual feature plays a major role in the identification of a multimodal meme since all the text-only baselines performed well. Multimodal baselines showed contradictory trends: ConcatBow showed a gain in precision, while ConcatBert showed a decline. This trend points to the need for more data since a complex model like ConcatBert needs more data to tune its hyperparameters. Hence, including more training data would be one of the future direction. Moreover, because of data imbalances, our baselines could not generalize classes with the least training samples even after introducing class weights -- undersampling classes with the majority, and oversampling ones in minority. Hence, we need a better approach or algorithm which will improve the evaluation metric despite the data imbalance. 
We think that it is important to understand the effect of troll or not-troll meme which might result in an opinion manipulation, and it is hard to identify troll meme with opinion manipulation, but we hope that the TrollsWithOpinion dataset will facilitate research in multimodal troll meme classification.

% {\color{red}
% Data statements in Appindex}

\begin{acknowledgements}
This publication has emanated from research supported in part by a research grant from Science Foundation Ireland (SFI) under Grant Number SFI/12/RC/2289\_P2, co-funded by the European Regional Development Fund, as well as by the H2020 project Prêt-à-LLOD under Grant Agreement number 825182.
\end{acknowledgements}

% BibTeX users please use one of
\bibliographystyle{spbasic}      % basic style, author-year citations
\bibliography{ref}   % name your BibTeX data base

% Non-BibTeX users please use

\end{document}